\documentclass[aps,prl,twocolumn,groupedaddress,longbibliography,nobibnotes,showpacs]{revtex4-1}
\usepackage{CJK}
\usepackage[english]{babel}

\usepackage{amsfonts,amsmath,amssymb,amsthm}
\usepackage{amsfonts,amsmath,amssymb,amsthm}
\usepackage{graphicx}
\usepackage{floatrow}
\usepackage[font=small,caption=false ]{subfig}
\newcommand{\fref}[1]{Fig.~\ref{#1}}
\newcommand{\eref}[1]{Eq.~\ref{#1}}
\newcommand{\bvec}[1]{\boldsymbol{#1}}

\newcommand{\rproj}[0]{R_{\text{proj}}}
\newcommand{\pswim}[0]{\Pi}
\newcommand{\teff}[0]{T_{\text{eff}}}
\newcommand{\teffstatic}[0]{T_{\text{eff,static}}}
\newcommand{\vrms}[0]{v_{\text{rms}}}
\usepackage[dvipsnames]{xcolor}
\newcommand{\NEW}[1]{\textnormal{#1}} %%submit
\begin{document}
\title{Testing a thermodynamic approach to collective animal behavior in laboratory fish schools} 
\author{Julia A. Giannini}
%\thanks{}
%%\affiliation{Department of Physics and Soft Matter Program, Syracuse University, Syracuse, New York 13244, USA}
\affiliation{Department of Physics, Syracuse University, Syracuse, New York 13244, USA}

\author{James G. Puckett}
\email[]{jpuckett@gettysburg.edu}
%\thanks{}
\affiliation{Department of Physics, Gettysburg College, Gettysburg, Pennsylvania 17325, USA}
\date{\today}

%==========================
%pacs; 3 max; descending relevance
%\pacs{87.50.yg}{Biophysical mechanisms of interaction}
%\pacs{87.23.Ge}{Dynamics of social systems}
%\pacs{05.70.Ce}{Thermodynamic functions and equations of state}
%\pacs{05.65.+b}{Self-organized systems}
%87.50.yg 	Biophysical mechanisms of interaction 
%87.23.Cc 	Population dynamics and ecological pattern formation 
%87.23.Ge 	Dynamics of social systems 
%05.65.+b 	Self-organized systems 
%%87.23.Cc, 05.65.+b, 87.23.Ge, 87.50.yg
%%87.23.Cc, 05.70.Ce, 87.23.Ge, 87.50.Y
%%87.50.cf Biophysical mechanisms of interaction 
%==========================
\begin{abstract}
Collective behaviors displayed by groups of social animals are observed frequently in nature.
Understanding and predicting the behavior of complex biological systems is dependent on developing effective descriptions and models. While collective animal systems are characteristically nonequilibrium, we can employ concepts from equilibrium statistical mechanics to motivate the measurement of material-like properties in laboratory animal aggregates. Here, we present results from a new set of experiments that utilize high speed footage of two-dimensional schooling events, particle tracking, and projected static and dynamic light fields to observe and control the behavior of negatively phototaxic fish schools (\textit{Hemigrammus bleheri}). First, we use static light fields consisting of dark circular regions to produce visual stimuli that confine the schools to a range of areas.  We find that schools have a maximum density which is independent of group size, and that a swim pressure-like quantity, $\Pi$ increases linearly with number density, suggesting that unperturbed schools exist on an isotherm.
Next, we use dynamic light fields where the radius of the dark region shrinks linearly with time to compress the schools. 
We find that an effective temperature parameter depends on the compression time and our results are thus consistent with the school having a constant heat flux.
These findings further evidence the utility of effective thermodynamic descriptions of nonequilibrium systems in collective animal behavior.

\end{abstract}

\maketitle

%\url{https://physics.aps.org/articles/v10/78#c7}
\section{Introduction}

Collective behavior is exhibited across many different scales in animal groups from microscopic bacteria or algae colonies \cite{Zhang2010Collective, Cates2012Diffusive,Marchetti2013Hydrodynamics}  to macroscopic insects \cite{Buhl2006disorder,Kelley2013Emergent,Puckett2015Searching,Attanasi2014Information}, schools of fish \cite{Berdahl2013Emergent,Tunstrom2013Collective,Puckett2018Collective} and flocks of birds \cite{Ballerini2008Interaction,Ballerini2008Empirical,Bialek2012Statistical}.
Collective behavior arises from self-organizing interactions between individuals \cite{Couzin2002Collective,Sumpter2010Collective} and can give rise to complex emergent group behaviors which surpass an individual's ability in navigation \cite{Berdahl2013Emergent,Puckett2018Collective} and foraging success \cite{Pitcher1982Fish,Bazazi2012Vortex}.   
In nature, groups can exhibit several morphologies or ``states" of collective animal behavior from disordered swarms to ordered flocks and mills.  
While individual animals may exhibit a range of behaviors and personalities \cite{Herbert-Read2013Role,Jolles2017Consistent}, early work showed that these collective `states' can be effectively modeled by active self-propelled particles (SPP) following uniform behavioral rules \cite{Vicsek1995Novel,Couzin2002Collective}.  
Moreover, the simplest and perhaps the most studied model, the Vicsek model \cite{Vicsek1995Novel}, exhibits a continuous phase transition between disordered (swarm) and ordered (flock) states \cite{Gregoire2004Onset,Szabo2006Phase,Chate2008Collective,Aldana2009Phase,Vicsek2012Collective}.  
In experiments, transitions between ordered and disordered phases have also been observed as the density of individuals increases \cite{Buhl2006disorder,Tunstrom2013Collective}.

As collective animal systems share many analogous features to nonliving active matter systems e.g., granular rods \cite{Ramaswamy2003Active,Narayan2007LongLived} and self-propelled colloids \cite{Palacci2013Living}, recent work has sought to build a unified framework to model soft matter systems.

Thus far, the theoretical approach has largely focused on studying active brownian particles (ABPs) which provide an ideal system to first construct a nonequilibrium thermodynamics for active matter \cite{Marchetti2013Hydrodynamics,Ramaswamy2010Mechanics,Patch2017Kinetics}.
In a thermal system, the ideal gas law, $P = \rho kT$, relates the mechanical pressure of the system confining the gas with the number density and temperature, where $\rho$ is the number density, $k$ Boltzmann's constant, and $T$ is the equilibrium temperature. 
However, whether or not the concepts of state variables such as pressure and temperature can be directly applied to nonequilibrium systems is an open question \cite{Cugliandolo2011effective}.  
Remarkably, simulated ABPs and experimental colloids have been shown to obey an equation of state much like their equilibrium counterparts \cite{Yang2014Aggregation,Mallory2014Anomalous,Takatori2014Swim,Takatori2015Thermodynamics, Ginot2015Nonequilibrium}, though the equation of state appears to only hold for spherical particles with torque-free interactions \cite{Solon2015Pressure,Solon2015Pressurea,Fily2017Mechanical}.  
Furthermore, in a recent experiment, researchers found that mechanical pressure does not equilibrate in a system of polar discs with different packing fractions and is therefore not a state variable \cite{Junot2017Active}.  
However, the pressure in an active matter system is still a useful quantity \cite{Omar2020Microscopic}, as even in anisotropic systems, the bulk pressure (swim stress) is self-consistent with the surface pressure \cite{Yan2018Anisotropic}.

An alternate approach taken in recent experiments is to borrow concepts and techniques from statistical mechanics and the physics of materials to test the effectiveness of state variable-like descriptions directly in animal groups.  
Recent experiments on insect groups have detected a density dependent phase transition \cite{Buhl2006disorder}, measured aggregate surface tension and viscosity \cite{Mlot2012Dynamics}, and found that groups of fire ants can form viscoelastic shear-thinning materials \cite{Tennenbaum2016Mechanics}.
In bird flocks, experimental observations have found scale-free behavioral correlations, \cite{Cavagna2010Scalefree} a signature of a continuous phase transition, and have shown that flocks can be modeled using a maximum entropy approach \cite{Bialek2012Statistical,Cavagna2017Dynamic,Cavagna2018Physics}.
Experiments on flying insect swarms have measured the susceptibility \cite{Attanasi2014Collective}, observed linear response to external perturbations (sound) \cite{Ni2015Intrinsic}, found that swarms have a finite modulus \cite{Ni2016tensile}, and observed phase co-existence \cite{Sinhuber2017Phase}.
Recently, experiments on midges have shown that a swarm responding to variable light stimuli contracts along an isotherm \cite{Sinhuber2019Response}.

%=====================
% experimental	FIGURE 1
%===================== 
\begin{figure}[t]
{\centering \includegraphics[width=0.95\linewidth]{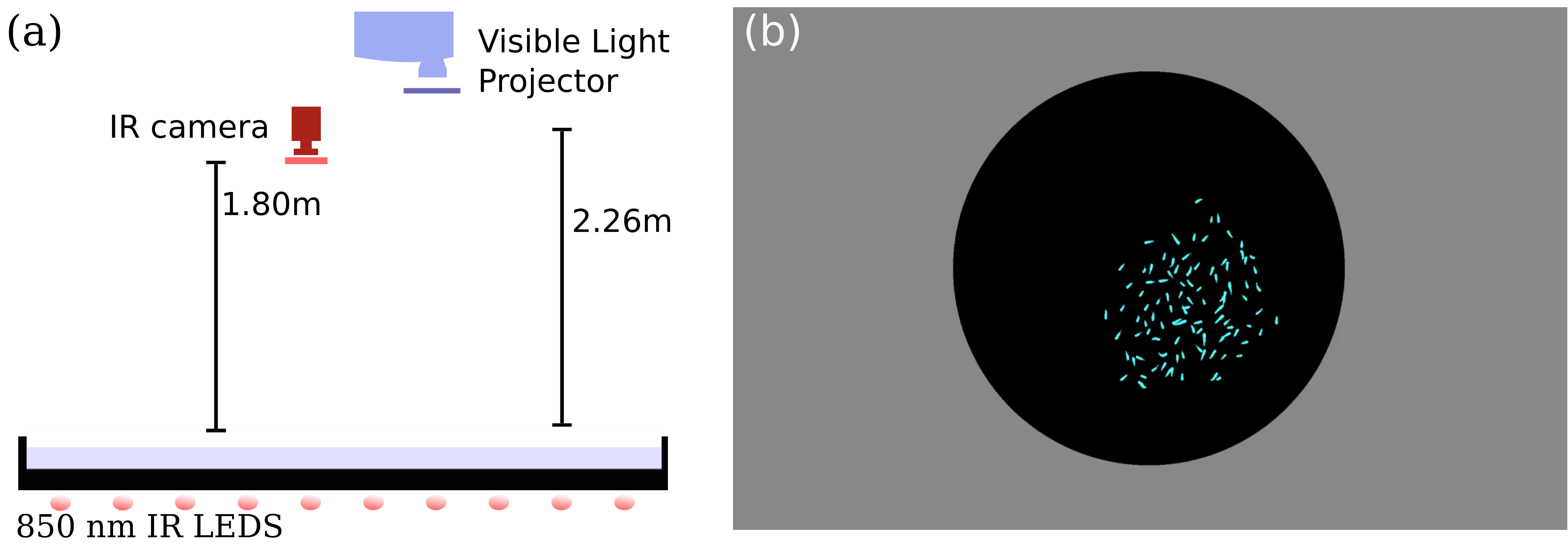}}
  \caption{(a) Schematic of our experimental setup. The experimental arena (a confined circular area 100 cm in diameter) lies inside a larger 122 by 244 cm tank.  Video footage is captured by a camera fitted with an infrared filter. 
 (b) Sample (static) visual stimulus used in experiments with the silhouettes of individual fish ($N=100$) overlaid for scale.}\label{fig:expt}
\end{figure}

In this work, we conduct a novel experiment aimed at investigating laboratory schools of rummy-nose tetra ({\it{Hemigrammus bleheri}}) in a thermodynamic framework. 
Since the tetra are negatively phototactic and prefer to be in dark regions in their environment, we project a circular dark disk on the center of a large quasi-two dimensional tank, as shown in \fref{fig:expt}(ab).  disk
By controlling the radius $\rproj$ of the projected dark spot, we confine the fish to a certain region of the tank as shown in \fref{fig:expt}(b).
Note that the fish are not mechanically confined, only weakly visually confined, and therefore can swim out of the central dark region with out experiencing any mechanical force.
We project both static and dynamic light fields and investigate the response of the fish to the different perturbations.
In the static light field experiments, the radius of the projected dark disk $\rproj$ is constant during each experimental trial.
In the dynamic light field experiments, we start with a radius $\rproj$ much larger than the unperturbed radii of the schools and reduce the radius of the projected dark disk with time.  
We then investigate the group level kinematic statistics of the schools and compare static and dynamic light perturbations.

%=====================
% Methods
%===================== 
\section{Methods and Results}

%=====================
% method
% Rproj = 56  89 130 179 236 300													in px
% Rproj = 49.33920705  78.41409692 114.53744493 157.7092511  207.92951542 264.31718062 	in cm
% Rproj = 1.64464023 2.61380323 3.81791483 5.25697504 6.93098385 8.81057269 			in BL
%
%===================== 

We filmed schooling events of groups of tetras (body length, BL=$3.4 \pm0.5$ cm) in a circular experimental arena (29.4 BL in diameter, 6~cm water depth, $T = 25\pm0.5^\circ C$). 
The arena (a clear acrylic wall in a circular shape) lies inside a larger 36BL by 72BL outer rectangular tank which houses water filtration and heating elements. %tank is 16.67BL
The outer tank provides a constant temperature heat bath to maintain the temperature of the inner circular tank.  
The projected visual stimulus extends beyond the inner tank and covers the entire outer tank, as fish can see through the clear inner wall. Fish were randomly selected from one of two home tanks (each tank having $N_{\text{home}} = 150$) and transferred to the experimental arena. 
Experiments were scheduled such that no individual was used in an experiment on two consecutive days. 
Before data collection, fish were given one hour to acclimatize to the experimental arena.

%=====================
% static		FIGURE 2
%===================== 
\begin{figure*}[t!]
\includegraphics[width=0.95\linewidth]{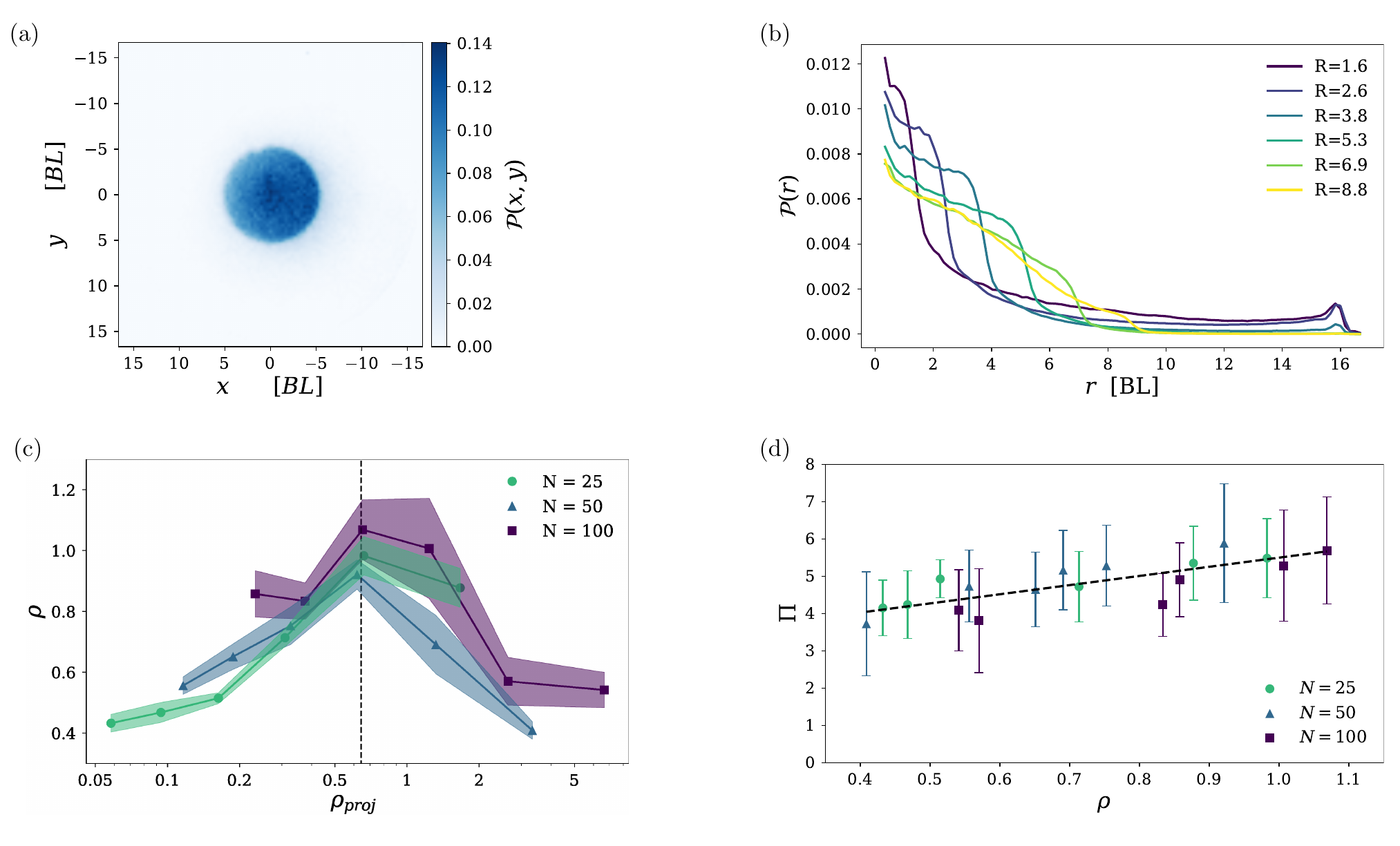}
  \caption{(a) The two-dimensional probability distribution for schools of $N = 100$ tetras and with a constant projected radius of $\rproj = 5.3$ BL. 
Depicts probability of finding an individual in a given region of the experimental arena.   
(b) Radial probability distributions for $N=100$ and $\rproj = 1.6, 2.6, 3.8, 5.3, 6.9, 8.8$ BL.
(c) Number density $\rho$ of schools as a function of the ratio of the group size $N$ and the volume of water within the projected dark disk for groups of $N = 25, 50, 100$ individuals. 
The maximum measured number density occurs at \NEW{$\rho_{\text{proj}} = 0.64$}, and is marked by the dashed vertical line.
(d) \NEW{The swim pressure-like $\pswim$} is shown as a function of average number density $\rho$ in trials with varying static $R_{\text{proj}}$ for group sizes of $N = 25, 50, 100$. 
The fit line represents an isotherm with an effective temperature for the static experiments of $kT_\text{static} \approx 2.45$. 
}
  \label{fig2}
\end{figure*}
%Figure 02(c) in Volume
%x_min   = 0.643818
%rho_max = 0.990929
%Figure 02(c) in Area
%x_min    = 1.136149
%rho_min = 1.538464

The fish schools were observed using an IR camera (PointGrey GS3-U3-41C6NIR-C) placed 180~cm over the tank, which records images at 4Mpx and  30~Hz, as shown in \fref{fig:expt}(a).
The tank was illuminated from below with an array of IR LEDs, which are invisible to the fish but visible to our camera.
Further details on video analysis, individual fish tracking, and information on fish husbandry are reported in a previous work \cite{Puckett2018Collective}.

As shown in \fref{fig:expt}(ab), a projector located 226~cm over the experimental arena casts a light field at 30~Hz onto the bottom of the tank. 
Throughout this work, the light field consist of a black disk (10 lux) with radius $\rproj$ on a grey background (150 lux).
A cropped sample image of the light field is shown in \fref{fig:expt}(b), with an overlay of the silhouette of a school of tetras for scale.

%=====================
% results 		STATIC
%===================== 
\subsection{Static experiments}

In our first series of experiments, the projected light fields consisted of a black disk with a constant radius $\rproj$.
We investigate the effect of confinement by the visual field as a function of six different static $\rproj = 1.65, 2.61, 3.82, 5.26, 8.81$~BL for three group sizes $N = 25, 50$ and 100.
For each trial, the static field was displayed for 1 min to allow the fish to reach a steady state, after which the camera recorded a video for 2 minutes.
We repeated experimental trials with randomly selected $\rproj$ and $N$ to generate 10 replicates for each $\rproj$ and $N$ combination.

%=====================
% results 		STATIC 		P(r) 			fig2ab
%===================== 
We found that the visual field provides an effective confinement as shown in \fref{fig2}(a), where the two-dimensional probability distribution indicates that the time-averaged structure of the school is axisymmetric. 	
The radial probability distributions ${\cal P}(r)$ for different $\rproj$ are shown for $N=100$ in \fref{fig2}(b).  
For the largest $\rproj$, we find that the experimental schools are entirely contained within $\rproj$.
However, for smaller $\rproj$, individuals may freely swim beyond $\rproj$, though
as expected, fish are more likely to be found near the center of the dark spot.
As shown in \fref{fig2}(b), we find ${\cal P}(r)$ decreases with $r$, quickly for small $\rproj$ and more gradually and approximately linearly for larger $\rproj$. 

In \fref{fig2}(c), we show that the light field provides a weak confinement which competes with overcrowding, as the fish do not exceed a maximal density.
The projected dark area weakly confines the fish and is our control parameter, where the ratio of the number of fish to the projected dark area is $\rho_\text{proj} = N/ (A_\text{proj} h_\text{water})$, where $h_\text{water} \approx 1.75$ BL.
The school changes size based on the projected dark area, where individual fish must balance an effective attraction to the dark area and an effective repulsion from overcrowding.
The measured number density of the school $\rho$ is determined by finding the quasi-two-dimensional area $A_\text{school}$ of the school and multiplying it by the water depth all in units of average body length. 
The area of the school is the area of the covariance error ellipse determining the 95\% confidence interval, which was found to be less susceptible to the locations of outlying fish compared to a simple convex hull area.
When $\rproj$ is large enough so that $\rho_{\text{proj}} \lesssim 0.5$, we see that the school adjusts its density roughly independent of $N$.
We also find that the school has a maximum number density per fish $\rho_{\text{max}} \approx 1.1$ fish/BL$^3$ to which the school will compress.  
At smaller $\rproj$ (large $\rho_\text{proj}$), large schools ($N=100$) extend beyond $\rproj$, where fish outside the projected dark disk are more disperse, decreasing $\rho$. Using the visual light field, we can control the the number density of the school $\rho$ by approximately a factor of two.

%=====================
% adapt for Referee B
%===================== 

%=====================
% results 		STATIC 		Pressure 		fig2d
%===================== 
Since the fish are visually (and weakly) confined within the projected light field, one cannot calculate a mechanical pressure since there is no momentum exchange between individuals and a container.  
The pressure for an ideal active matter system can be derived as $\Pi_\text{swim} = n \zeta U_0^2 \tau_R /2$ for two dimensional systems\cite{Takatori2014Swim}, where $\zeta$ is the hydrodynamic drag and $U_0 \tau_R$ is the run length in a reorientation time.  
However, in many experiments, either due to finite size effects or insufficiently length of trajectories, these terms are not measurable.   
Similar quantities to the swim pressure have been derived from a virial equation \cite{Gorbonos2016Longrange} and were shown to relate to thermodynamic phases in other collective animal systems~\cite{Sinhuber2017Phase, Sinhuber2019Response}, even though these ``pressures" are not equivalent to the mechanical pressure.

We define a ``pressure" similar to previous experiments, where the swim pressure-like quantity (per unit mass) is related to the ratio of kinetic energy per unit volume,

\begin{equation}
  \begin{array}{l}
    \pswim =    \left\langle \dfrac{N}{V} \cdot \dfrac{1}{N} \sum\limits_{i=1}^N  \dfrac{1}{2} \bvec{v}_i^2 \right\rangle_t
  \end{array}
  \label{eqn1}  
\end{equation}
where $N$ is the number of individuals, $V$ is the volume occupied by the school, $\bvec{v}_i$ is the velocity of individual of fish $i$, and the notation $\langle \rangle_t$ denotes a time-average.
Since the system is quasi-two-dimensional, V is measured by $A_\text{school} h_\text{water}$.
This quantity is the average ratio of the total kinetic energy to the volume of the school.
Written more compactly below in terms of number density $\rho=N/V$ and the rms velocity, we have,
\begin{equation}
  \begin{array}{l}
    \pswim = \left\langle\rho \vrms^2 \right\rangle_t,
  \end{array}
  \label{eqn2}
\end{equation}
where both $\rho$ and $v_{\text{rms}}$ are functions of time. 
This equation is analogous to the classical definition  $P = \rho kT$, where $\rho=N/V$, $k$ is Boltzmann's constant, and $T$ is the temperature. 
While $\Pi$ and $\vrms$ are positively correlated, both $\rho$ and $\vrms$ are time dependent quantities, so $\Pi$ can not be interchanged with or depend solely on $\vrms$.

In \fref{fig2}(d), we show that $\pswim$ increases linearly as function of $\rho$ ($F_{1,16}=27.15$, $p=8.6\times10^{-5}$) for schools of different $N$ and $\rproj$.
The slope of $\pswim(\rho)$ yields an effective temperature of $k \teff = 2.45$ for our set of static experiments.
While this result may at first glance appear trivial, that the fish swim at roughly the same speed independent of the density or group size (e.g., $\vrms^2 = $ const.), this is not the case as our subsequent dynamic experiments show.

%=====================
% results 		DYNAMIC
%===================== 
\subsection{Dynamic experiments}

\begin{figure}[t!]
\centering \includegraphics[width=0.95\linewidth]{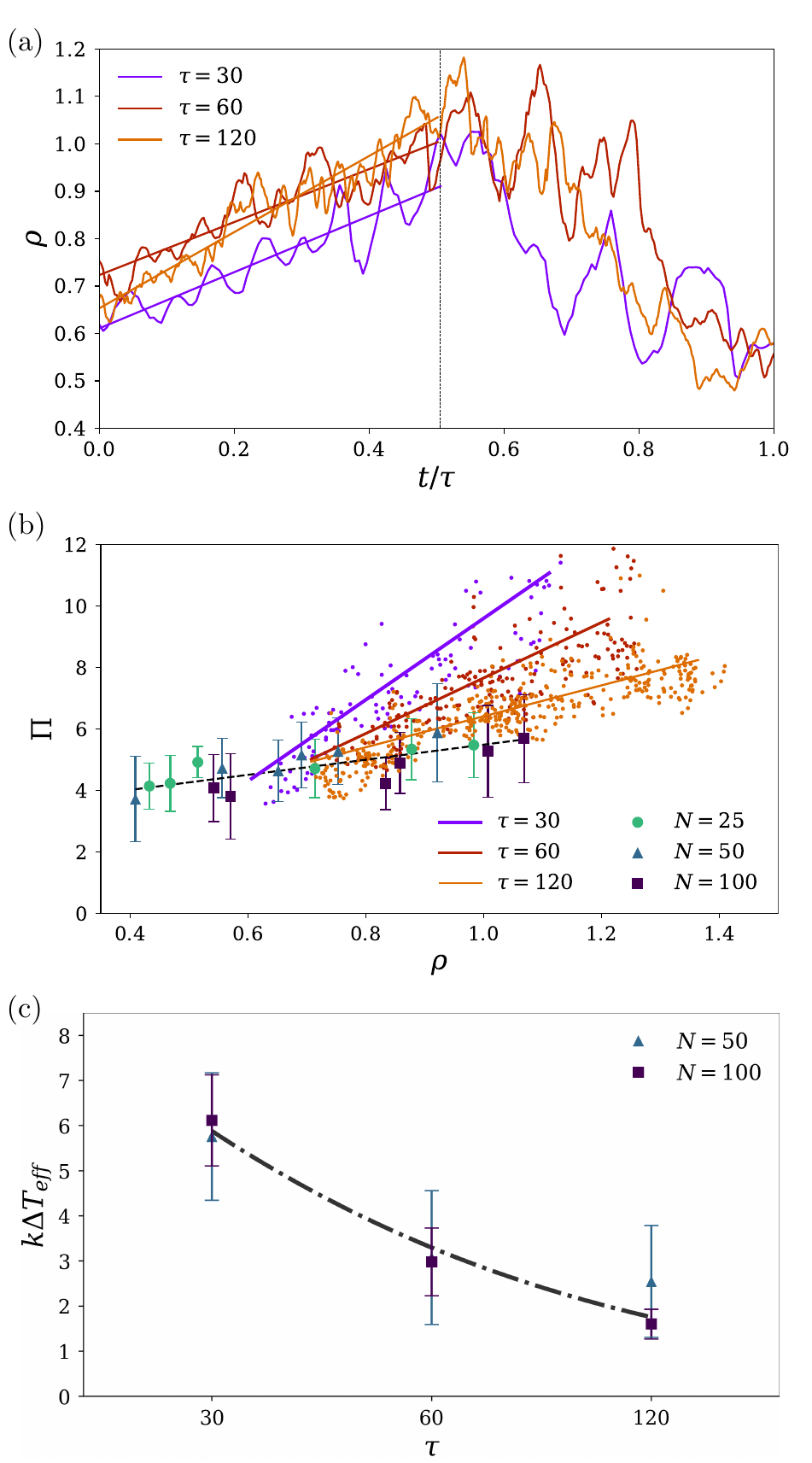}

  \caption{(a) The number density of $N = 100$ tetra schools as a function of time normalized by $\tau$ for $\tau = 30, 60, \text{ and } 120$ s. We see that the school is compressible until it reaches the maximum density and afterward decreases in density and shoals freely. 
(b) \NEW{The swim pressure-like quantity $\pswim$} plotted as a function of number density for group size $N = 100$ and $\tau = 30, 60$ and $100$ s. For comparison, this data is overlaid on the data for the static experiments. The effective temperature of the school is computed via a least squares linear fit. 
(c) Average change in effective temperature of the system between the dynamic and static systems as a function of compression time $\tau$.  The fit line is a nonlinear least squares fit to \eref{eqn5}.}
  \label{fig3}
\end{figure}

In our second series of experiments, projected light fields consisted of a light background and dark disks, where the radius of the disk, $\rproj(t) = R_\text{0} \left( 1 - t/\tau \right)$, decreases linearly in time from $R_0=14.1$~BL to 0 in $\tau$ seconds.
We conducted trials with $\tau = 120, 60$, and $30$~s and for group sizes $N = 50$ and $100$, each with 10 replicates.  
%%We did not conduct dynamic experiments on $N=25$ due to poor statistics.  let the ref point that out.
%%\jag{(not 10? Should we mention that $N = 25$ is not super effective?)}.  I don't think we need to
In \fref{fig3}(a), we show the number density $\rho$ of the school as a function of time with $N=100$ for each $\tau$, where time is rescaled by $\tau$. 
We find that the number density $\rho$ of the schools increases roughly linearly with time as $\rproj$ decreases, until about halfway through the trial, which corresponds to the approximately the same maximum number density $\rho_\text{max}$ noted in static experiments.
After the maximum number density is reached, the area of the school grows and subsequently begins to shoal freely as the area of the dark disk is much smaller than the area of the school.
Since the school is not further compressed after $\tau / 2$, we do not use this data for any subsequent results.

%=====================
% dynamics		FIGURE 3
%===================== 

In \fref{fig3}(b), we compute pressure-like $\pswim$ as a function of $\rho$ with $N=100$ for each compression time $\tau = 30,~60$ and $100$.
Note that $\pswim$ here is no longer time averaged, since $\rho$ changes with time. 
Even for a slow compression $\tau=120$s, the pressure $\pswim$ increases faster as a function of $\rho$ compared to the static case. 
Therefore, the effective temperature $\teff$ (the slope of the isotherm on $\pswim(\rho)$) increases as the compression times $\tau$ decrease.
The school not only adjusts its number density $\rho$ in response to the shrinking projected disk, but also changes its kinetic energy ($\frac{1}{2} \vrms^2$) based on the rate of change of the dark area.
When the system is compressed, the size of the school decreases but there is also a change in the velocities of the fish, where shorter compression times $\tau$ result in larger increases in kinetic energy.

%=====================
% results 		proof
%===================== 

In \fref{fig3}(c), we show the change in effective temperature $k \Delta \teff$ as a function of $\tau$ for different group sizes $N$, where $k \Delta \teff = k\teff - k\teffstatic$.  
Here, we find that $k \Delta \teff$ increases with decreasing $\tau$.

Although the school is far from equilibrium, for a first attempt to understand the relationship between $k \Delta \teff$ and $\tau$, we begin with borrowing ideas from classical equilibrium statistical mechanics.  The first law of thermodynamics relates the heat added to the system $Q$ and the work done on the system $W$ to the change in internal energy 
 \begin{align}  	
   \Delta U = Q+W
  \label{eqn3}
  \end{align}
where the change in internal energy is proportional to the change in temperature, giving $\Delta U =   \alpha \Delta \teff$, where $\alpha$ is a constant.

Since all dynamic experiments start with the same $\rho_\text{initial}$ and end with $\rho_\text{final}$, they all do the same amount of work. 
Furthermore, to reach the density $\rho_\text{final} \approx \rho_\text{max}$, it takes approximately half the compression time, $\frac{ \tau }{2}$, as shown in \fref{fig3}(a).  
While this work, like the change in internal energy is extensive, for the sake of simplicity of the model, we ignore the dependence of each on $N$, and instead concentrate on the time dependence of heat loss.

Given a finite power of heat loss in the system, a thermodynamic process doing the same amount of work in less time should increase the temperature of the system.  Therefore, the heat lost by the system as it is compressed to twice the initial density is $Q =\overline{P}_\text{heat}  \frac{ \tau }{2}$, where the  rate of heat loss $\overline{P}_\text{heat}$ is proportional to the temperature difference $\Delta \teff$.
We can rewrite $Q = -\beta \tau \Delta \teff $.

Using each of these relationships, we can rewrite \eref{eqn3} as
 \begin{align}  	
   \alpha  \Delta  \teff &= W - \beta  \tau  \Delta  \teff .
  \label{eqn4}
  \end{align}
 Solving for $\Delta \teff$, we have
\begin{equation}
  \begin{array}{l}
    \Delta \teff = \dfrac{W}{ \alpha + \beta  \tau} = \dfrac{1}{a+b\tau}
  \end{array}
  \label{eqn5}
\end{equation}

There are then two fit parameters, $a = \alpha/W$ and $b = \beta/W$, for 57 dynamic runs for both $N=50$ and $100$ and $\tau=30$, $60$, and $120$s. Using nonlinear least squares fitting on all the dynamic data ($N=50$ and $100$), we find $a = 0.037 \pm 0.063$ and $b=0.0044\pm 0.0018$ s$^{-1}$.
%57 = 3*2*
%57= 2*3* 9.5;  
%b: 0.004438707531573303+/-0.001792020236144575
%a:  0.03689944558659977 +/-0.06306063479474973
%

While the derivation of \eref{eqn5} was made with numerous idealized assumptions, remarkably, we find the uncertainty in $b$ is small compared to its value, and the model is in good agreement with the experimental data.  
Certainly, there are several ways to treat this derivation more rigorously including: considerations of $N$ dependence for $\Delta U$, $Q$ and $W$, or accounting for the rate of work done during the compression in the expression for the rate of heat loss.
However, given this simplified and arguably na\"ive treatment, our results are consistent with the argument for a `finite heat flux' of the school, where a faster dynamic compression yields a larger change in effective temperature.
%%b = 0.00783 +/- 0.00142 		%heat flux
%%a = 0.06512 +/- 0.04995

\section{Conclusions and Future Work}

%discussion
Our findings show that rummy-nose tetra ({\it{Hemigrammus bleheri}}) can be visually confined with projected light fields and that their number density $\rho$ can be controlled without applying any physical force. 
Using static light fields, we find that $\pswim$ depends linearly on the number density $\rho$ consistent with the behavior of an isotherm, and that schools exhibit a common effective temperature that is independent of the degree of static confinement imposed on the system.
In response to dynamic light fields, the school adjust its number density $\rho$ to fit within the projected dark area, balancing the favor of being in the dark area with overcrowding.
While $\pswim$ and $\rho$ are still linearly related, we find that the slope or effective temperature is dependent on the compression time, with faster compression leading to larger $T_\text{eff}$. 
Note that while $\rho_\text{proj}$ grows quadratically with time, we find that $\rho$ increases approximately linearly with $t/\tau$.  
This may indicate that $\rho$ is rate limited by biology or physiology. 
Additionally, fish on the boundary may respond more strongly to the shrinking dark disk, making $\rho$ nonuniform and larger at the boundary.

%%%%%%
%needs work
%%%%%%
Furthermore, our dynamic studies show that there is a finite rate of heat lost during these dynamic compressions.  This heat loss drives the system back to $\teffstatic$ likely because the fish have a preferred swim speed. 
As stated above, in the experiments utilizing static light fields, the effective temperature remained constant for all confining areas. 
We propose that this is due to the fish consistently achieving their preferred swim speed in the static experiments. 
This observation is consistent with  \eref{eqn5} given $\tau \rightarrow \infty$, where an infinitely slow compression should yield the same effective temperature observed in the static experiments.
However, for the dynamic compression experiments, the compression time $\tau$ determines both the number density $\rho(t)$ and the swim speed $v_\text{rms}(t)$. 
For decreasing $\tau$, the effective temperature increases as work is done but there is less time for heat to dissipate.  
Many possible biological or physiological reasons exist for this finite rate of heat loss which may include finite response times to either the visual stimuli (mainly concerning fish on the perimeter) or inter-individual kinetics.

Future work may explore various extensions including investigating larger schools or a wider range of $\tau$.
In particular, one could explore new thermodynamic processes on schools of social fish (e.g., effective adiabatic compression) or extend these methods to other social animals.
In this regard, investigating the response of group behavior to perturbations may lead to the construction of proper definitions for pressure-like and temperature-like variables for collective animal behavior and other active matter systems.

\vspace{0.5em}

\begin{acknowledgments}
We thank N. Ouellette and N. Gov for illuminating discussions.   
This work is also supported by Gettysburg College and by the Cross-Disciplinary Science Institute at Gettysburg College (X-SIG).
\end{acknowledgments}

%==================================
%==================================
%\bibliography{2020thermo}
%==================================
%==================================
%

%==================================
%==================================
\end{document}